\documentclass[reprint, aps, prd, amsmath, amssymb, nofootinbib]{revtex4-2}

\usepackage{graphicx} 
\usepackage{dcolumn}  
\usepackage{bm}       

\usepackage[colorlinks=true
,citecolor=blue
]{hyperref}
\usepackage{lipsum}
\usepackage{siunitx}
\DeclareSIUnit\year{yr}
\usepackage{enumitem} 
\usepackage{soul}
\usepackage{booktabs}
\usepackage{slashed}

\graphicspath{{img/}} 

\usepackage{graphicx}

\setlist[description]{font=\normalfont\space}

\DeclareRobustCommand{\rchi}{{\mathpalette\irchi\relax}}
\newcommand{\irchi}[2]{\raisebox{\depth}{$#1\chi$}} 

\begin{document}

\title{Dimension-5 baryon-number violation in low-scale Pati--Salam}

\author{Tomasz P. Dutka}
\email{tdutka@kias.re.kr}
\affiliation{School of Physics, Korea Institute for Advanced Study, Seoul 02455, Republic of Korea}

\author{John Gargalionis}
\email{john.gargalionis@ific.uv.es}
\affiliation{Departamento de F\'isica Te\'orica and IFIC, Universidad de
  Valencia-CSIC, C/Catedr\'atico Jos\'e Beltr\'an, 2, E-46980 Paterna, Spain}

\begin{abstract}
  The gauge bosons of Pati--Salam do not mediate proton decay at the renormalisable level, and
  for this reason it is possible to construct scenarios in which
  $\mathrm{SU}(4) \otimes \mathrm{SU}(2)_{R}$ is broken at relatively low scales. In this paper we show
  that such low-scale models generate  dimension-5 operators that can give rise to nucleon decays at unacceptably large rates, even if the operators are suppressed by the Planck scale. We find an interesting complementarity between the nucleon-decay limits and the usual meson-decay constraints. Furthermore, we argue that these operators are generically present when the model is embedded into $\mathrm{SO}(10)$, lowering the suppression scale.
  Under reasonable assumptions, the lower limit on the breaking scale can be constrained to be as high as $\mathcal{O} (10^{8})$~GeV.
\end{abstract}

 \maketitle

\section{Introduction}

The Pati--Salam (PS) model~\cite{Pati:1973rp,Pati:1973uk,Pati:1974yy} is a compelling framework for quark--lepton unification and potentially a stepping stone towards grand unification. Theories which contain any form of quark--lepton unification are usually expected to suffer from stringent proton decay limits on their symmetry breaking scales. However, like the Standard Model (SM), the theory contains an accidental baryon-number symmetry preventing such processes. This allows limits on the PS breaking scale to be set by the non-observation of flavour-changing neutral currents (FCNCs) rather than bounds on the stability of nucleons. These FCNC bounds are significantly lower than the scale implied by Grand Unified Theories (GUTs). As such there is the tantalising prospect that Pati--Salam, or a similar theory, can be experimentally tested and verified.

Investigations of PS scenarios with low-scale breaking date to the original paper~\cite{Pati:1974yy}, but have seen a resurgence of interest recently following indirect evidence of the leptoquark $\mathrm{SU}(4)$ gauge boson in the context of explanations of the evolving neutral- and charged-current flavour anomalies, see \textit{e.g.}~\cite{Alonso:2014csa,Calibbi:2015kma,Barbieri:2015yvd,Fajfer:2015ycq,Bhattacharya:2016mcc,Buttazzo:2017ixm,Angelescu:2018tyl,Angelescu:2021lln,Aebischer:2022oqe,London:2021lfn}. Limits imposed from the experimental absence of FCNCs generally require that the PS breaking scale sit higher than $ \mathcal{O}(10^3)$ TeV. Careful choices of parameters can act to evade these bounds by an order of magnitude~\cite{Kuznetsov:1994tt,Valencia:1994cj,Kuznetsov:1995wb,Smirnov:2007hv,Smirnov:2008zzb,Smirnov:2018ske,Dolan:2020doe} with specific flavour structures of mixing matrices. Generally, modifications of the theory must be made (see \textit{e.g.}~\cite{Balaji:2018zna,Balaji:2019kwe,Cornella:2019hct}) to allow the ultra-low-scale breaking required to accommodate the leptoquark at a few TeV, the scale suggested by global fits of SMEFT coefficients to $b \to s$ and $b \to c$ measurements, see \textit{e.g.}~\cite{Aebischer:2019mlg}, or in order for collider probes to become relevant in such theories.

In this paper we propose that a minimal set of ingredients generally present in low-scale $\mathrm{SU}(4) \otimes \mathrm{SU}(2)_{L} \otimes \mathrm{SU}(2)_{R}$ models is sufficient to induce the accidental violation of baryon number through dimension-5 effective operators. These operators mediate nucleon decays at dangerously large rates, placing bounds on the PS breaking scale up to two orders of magnitude higher than those implied by lepton-flavour-violating $K_L^0$ decays, depending on some assumptions. Importantly, as we will show, the flavour structures required in order to suppress $K_L^0$ decays do not align with those needed to suppress these $B$-violating nucleon decays. The most model-independent bound follows by turning on the baryon-number violation only at the Planck scale, and the same argument has been made for scalar leptoquarks more generally, \textit{e.g.}~\cite{Arnold:2013cva,Assad:2017iib,Herrero-Garcia:2019czj,Murgui:2021bdy,Davighi:2022qgb}. Furthermore, we argue that these effects should in fact be present already at the scale of $\mathrm{SO}(10)$ unification from an analysis of the tree-level completions of the dimension-5 operators. This leads to a mild enhancement of the nucleon decay rates, allowing the breaking scale to be constrained to as high as $\mathcal{O}(10^8)$~GeV.

\section{Minimal setup}

In this section we outline the ingredients generically present in those PS
models that allow $\mathrm{SU}(4)$ breaking at scales as low as
phenomenologically possible. We try to be maximally agnostic with regards to
details of the model that are not directly relevant to the focus of this study:
accidental baryon-number violation in the effective Lagrangian. This includes
most details concerning breaking the down-type quark and charged-lepton mass relations,
the presence of additional fermions in the theory, and the usual hierarchy problems\footnote{We note however that models with the particle content we will assume can be made supersymmetric, \textit{e.g.}~\cite{Antoniadis:1988cm}.}. We intend the following discussion to define our use of the term \textit{low-scale PS} in this paper.

In the canonical Pati--Salam model, the fermion content of the SM is extended by three right-handed neutrinos and is then unified
into the representations
\begin{equation}
  \label{eq:sm-fermions}
  \begin{split}
    (f_{L})_{p} &= \begin{pmatrix} Q_{L} & L_{L} \end{pmatrix}_{p} \sim (\mathbf{4}, \mathbf{2}, \mathbf{1})_{p} \ , \\
    (f_{R})_{p} &= \begin{pmatrix} u_{R} & \nu_{R} \\ d_{R} & e_{R} \end{pmatrix}_{p} \sim (\mathbf{4}, \mathbf{1}, \mathbf{2})_{p} \ ,
  \end{split}
\end{equation}
of the Pati--Salam gauge group
$G_{\text{PS}} \equiv \mathrm{SU}(4) \otimes \mathrm{SU}(2)_{L} \otimes \mathrm{SU}(2)_{R}$,
for each generation, \textit{i.e.} $p \in \{1,2,3\}$. The lowest-dimensional multiplet
that breaks the gauge symmetry to that of the SM is the scalar field $\chi$:
\begin{equation}
    \label{eq:ps-scalar}
  \chi = \begin{pmatrix} \chi^{u} & \chi^{0} \\ \chi^{d} & \chi^{-} \end{pmatrix} \sim (\mathbf{4}, \mathbf{1}, \mathbf{2}) \ ,
\end{equation}
through the vev of the neutral component: $v_R \equiv \sqrt{2} \langle \chi^{0} \rangle$. In this minimal setup $\chi^u$ is the Goldstone boson of the $\mathrm{SU}(4)$ vector-leptoquark gauge boson (from hereon referred to as $X_\mu$ with mass $m_X$) while $\chi^-$ and $\chi^0$ predominately make up the Goldstone bosons of the $W'_\mu$ and $Z'_\mu$ respectively~\cite{Volkas:1995yn}.

The mass of the physical leptoquark scalar $\chi^d$ results from minimising the scalar potential:
\begin{equation}
    m_{\chi^d}^2 \simeq -\frac{1}{2} \lambda^\rchi v_R^2 \ .
\end{equation}
where $\lambda^\rchi$ is the coupling of $(\chi^\dagger \chi)^2$. Limits on the value of $|\lambda^{\rchi}|$ can be derived from the unitarity of $\chi\chi$-scattering. Assuming only a contribution from the $\lambda^{\rchi}$ term and imposing the partial-wave unitarity bound $|\text{Re}(a_0)| \leq \frac{1}{2}$, we find $|\lambda^{\rchi}| \leq 2\pi$, and therefore
\begin{equation}
  \label{eq:mchid-vr-relation}
  m_{\chi^d} \leq \sqrt{\pi} v_{R} \ .
\end{equation}

The scalar field
$\Phi \sim (\mathbf{1}, \mathbf{2}, \mathbf{2})$:
\begin{equation}
  \Phi = \begin{pmatrix} \Phi_{1} & \Phi_{2} \end{pmatrix} =  \begin{pmatrix} \phi_{1}^{0} & \phi_{2}^{+} \\ \phi_{1}^{-} & \phi_{2}^{0} \end{pmatrix} \ ,
\end{equation}
is also introduced to break EW symmetry and generate masses for the
fermions of Eq.~\eqref{eq:sm-fermions}:
\begin{equation}
  \label{eq:ps-yukawas}
  \mathcal{L} \supset y^{pq} \mathrm{Tr}[(\bar{f}_{L})_{p} \Phi (f_{R})_{q}] + \tilde{y}^{pq} \mathrm{Tr}[(\bar{f}_{L})_{p} \tilde{\Phi} (f_{R})_{q}] + \text{H.c.} \ ,
\end{equation}
with $\tilde{\Phi} \equiv \tau_{2} \Phi^{*} \tau_{2}$. The breaking of EW
symmetry occurs through the vevs
\begin{equation}
  \langle \Phi \rangle = \frac{1}{\sqrt{2}}\begin{pmatrix} v_{1} & 0 \\ 0 & v_{2} \end{pmatrix} \ .
\end{equation}
Expanding Eq.~\eqref{eq:ps-yukawas} yields the relations between the singular values $\sigma_{p}$ of the mass matrices
\begin{equation}
  \sigma_p(\mathbf{M}_{d}) = \sigma_p(\mathbf{M}_{e}) \quad \text{ and } \quad \sigma_p(\mathbf{M}_{u}) = \sigma_p(\mathbf{M}_{\nu})
\end{equation}
at the scale of EW symmetry breaking. Here we define $\mathbf{M}_i$ to be the $3 \times 3$
mass-mixing matrix for the three generations of SM fermions of type $i$,
expected in the minimal model. The second of these relations is far more
grievous than the first and is most simply
corrected by introducing the Weyl state
$\xi \sim (\mathbf{1}, \mathbf{1}, \mathbf{1})$ to organise for light neutrino
masses.

An important property of Pati--Salam, relevant below when discussing $B$-violating nucleon-decay limits, is the existence of eight physical mixing matrices between the fermion generations which are generalisations of the CKM and PMNS matrices in the SM. Their definitions can be found in Ref.~\cite{Dolan:2020doe}, however an important identity that will be used later is the relationship between them:
\begin{equation}
\label{eqn:mixingmatrixconnection}
    \mathbf{K}_{L/R}^{u\nu} = \mathbf{V}_{\text{CKM}}^{L/R} \mathbf{K}_{L/R}^{d e} \mathbf{U}_{\text{PMNS}}^{L/R} \ .
\end{equation}
Here, $\mathbf{K}_{L}$ and $\mathbf{K}_{R}$ are mixing matrices between $X_\mu$, the physical up-quarks and neutrinos, as well as between the down-quarks and charged leptons for the left- and right-handed fields respectively.

\paragraph*{Neutrino sector}
In order to more clearly illustrate the nucleon-decay mechanism, we continue to derive the neutrino mixing and a representative nucleon decay bound in a one-generational model. We generalise our results to the three-generational framework in Sec.~\ref{sec:proton-decay}.
 
To reduce clutter, we also write $\overline{\psi_1^{c}} \psi_2$ as $\psi_1 \psi_2$ for
fermions $\psi_1$ and $\psi_2$, where $\psi_1^{c}$ is the usual charge conjugate of
$\psi_1$.

The field $\xi$ introduces Yukawa couplings for the scalar $\chi$, and it could have a Majorana mass $\mu$:
\begin{equation}
  \label{eq:xi-lag}
  \mathcal{L} \supset - \frac{\mu}{2} \xi \xi - y_{R} \bar{\xi} \textrm{Tr}[\chi^{\dagger} f_{R}] + \text{H.c.}
\end{equation}
Expanding out the second term using Eqns.~\eqref{eq:sm-fermions} and \eqref{eq:ps-scalar},
\begin{equation}
\label{eq:LQcouplingexpanded}
    \overline{\xi}\left[ (\chi^d)^\dagger d_R + (\chi^-)^\dagger e_R + (\chi^u)^\dagger u_R + (\chi^0)^\dagger \nu_R \right],
\end{equation}
it is simple to see that a mixing will be induced between $\xi$ and $\nu_R$ once $\chi^0$ obtains a vacuum expectation value. Therefore the first and third terms of Eq.~\eqref{eq:LQcouplingexpanded} can be interepted as genuine leptoquark couplings, albeit suppressed by mixing angles.

In the presence of these terms, the neutrino-mass matrix takes the form
\begin{equation}
    \begin{pmatrix} \bar{\nu}_{L} && \nu_{R} && \bar{\xi} \end{pmatrix} 
  \begin{pmatrix}
    0 & m_{u} & 0 \\
    m_{u} & 0 & y_{R} v_{R} \\
    0 & y_{R} v_{R} & \mu \\
  \end{pmatrix}
    \begin{pmatrix} \nu_{L}^c \\ \nu_{R} \\ \xi^c \end{pmatrix}\ .
\end{equation}
Diagonalisation gives rise to a pseudo-Dirac fermion, composed of $N_L$ and $N_R$, and a Majorana fermion $\nu$. Assuming the hierarchy
$\lvert  m_{u}, \mu \rvert \ll \lvert y_{R}v_{R} \rvert$, we find
\begin{equation}
  \label{eq:physical-neutrinos}
  \begin{split}
   N_{L} &\simeq \sin \theta \nu_{L} + \frac{m_u \mu}{(y_R v_R)^2} \nu_R^c+ \cos \theta \xi \ , \\
  N_{R} &\simeq -\frac{m_u \mu}{(y_R v_R)^2}\nu_L^c + \nu_{R} + \frac{1}{2}\frac{m_u^2 \mu}{(y_R v_R)^3} \xi^c\ , \\
  \nu &\simeq \cos \theta \nu_{L} + \frac{1}{2}\frac{m_u^2 \mu}{(y_R v_R)^3} \nu_R^c - \sin \theta \xi \ ,
  \end{split}
\end{equation}
up to $\mathcal{O}(\mu)$ with
\begin{equation}
\label{eq:one-gen-mixing}
\tan \theta = \frac{m_{u}}{y_{R} v_R} \ .
\end{equation}
The physical masses, at lowest order in $\mu$, are given by
\begin{align}
  m_{N_{L/R}} &\simeq \sqrt{\lvert y_{R} v_R\rvert^{2} + \lvert m_{u}\rvert^{2}} \pm \frac{1}{2}\mu \\
  m_\nu &\simeq \left(\frac{m_{u}}{y_{R} v_R}\right)^2 \mu.
\end{align}
Note that in the limit $\mu \rightarrow 0$, $\nu$ is an exactly massless Weyl fermion, while $N_L$ and $N_R$ form a genuine Dirac fermion. In fact, this low-scale setup was the one first proposed in the original iteration of the model~\cite{Pati:1974yy} to organise for exactly massless neutrinos with $\mu = 0$. With $\mu$ taking a small but non-zero value, $\nu$ develops a mass linearly
proportional to $\mu$ and can therefore be small in a technically
natural way without the need for large Majorana masses generated at high PS
breaking scales. 

The Lagrangian contains one global $\mathrm{U}(1)$ symmetry:
\begin{equation}
    \mathrm{U}(1)_{J}:\quad f_{L,R} \to e^{i\theta J} f_{L,R},\quad \chi \to e^{i\theta J} \chi \ .
\end{equation}
Baryon- and lepton-number can be identified with different linear combinations of $J$ and the diagonal generator of $\mathrm{SU}(4)$ that commutes with the unbroken $\mathrm{SU}(3)$ subgroup, which we call $T$. The normalisation of $T$ is chosen such that $Y = T_{3R} + T/2$, \textit{i.e.} such that $T \equiv B - L$. We find
\begin{equation}
B = \frac{1}{4} (J + T ),\quad L = \frac{1}{4} (J - 3 T) \ .
\end{equation}

\section{Effective Lagrangian}

\begin{table}[t]
  \centering
  \bgroup
   \def\arraystretch{1.3}
    \begin{tabular}{ccc}
    \toprule
    Field content $\quad$ & $J$  & \quad Number of operators \\
    \midrule
    $\Phi^\dagger \Phi^\dagger \xi \xi$ & $0$ &  $n_{\xi} (n_{\xi} + 1) / 2$ \\
    $\Phi^\dagger \chi^\dagger f_L \xi$   &$0$&  $n_{\xi} n_f$ \\
    $\Phi^\dagger \Phi \xi \xi$ &  $0$ & $n_{\xi} (n_{\xi} + 1) / 2$ \\
    $\chi^\dagger \chi^\dagger f_L f_L$ & $0$ &  $n_{f} (n_{f} + 1) / 2$ \\
    $\chi^\dagger \chi^\dagger f_R f_R$ & $0$ &$n_{f} (n_{f} + 1)$ \\
    $\Phi \chi^\dagger f_L \xi$  & $0$ &$n_{\xi} n_{f}$ \\
    $\chi^\dagger \chi \xi \xi$  & $0$ & $n_{\xi} (n_{\xi} + 1) / 2$ \\
    $\chi \chi f_L f_L$  & $4$ &$n_{f} (n_{f} + 1) / 2$ \\
    $\chi \chi f_R f_R$  & $4$ & $n_{f} (n_{f} + 1) / 2$ \\
    $\Phi \Phi \xi \xi$  & $0$ &$n_{\xi} (n_{\xi} + 1) / 2$ \\
    \bottomrule
  \end{tabular}
  \egroup
  \caption{Combinations of fields appearing at dimension-5 in the effective Lagrangian of our minimal setup, along with their net global-symmetry assignment and the operator counting for $n_f$ SM-fermion generations and $n_{\xi}$ generations of $\xi$.
  }
  \label{tab:eff-ops}
\end{table}

There are 10 combinations of fields that constitute the effective Lagrangian at
dimension 5. These are presented in Table~\ref{tab:eff-ops} along with their net violation of $J$ as well as their permutation-symmetry properties. The operators of interest to us are
those that violate baryon and lepton number, and we find these to be $\chi^2 f_X^2$. 
Concretely, we define
\begin{equation}
  \label{eq:b-viol-operators}
  (\mathcal{O}_{X})_{pq} \equiv (f_{X})^{\alpha i}_{p} (f_{X})^{\beta j}_{q} \chi^{\gamma k} \chi^{\delta l} \epsilon_{\alpha \beta \gamma \delta} \epsilon_{ij} \epsilon_{kl} \ ,
\end{equation}
where $X \in \{L, R\}$, Greek letters $\alpha,\beta,\gamma,\delta$ are $\mathrm{SU}(4)$ fundamental indices, and $i,j,k,l$ are the relevant $\mathrm{SU}(2)$ fundamental indices. The coefficients are normalised such that
\begin{equation}
  \label{eq:dim-5-lag}
  \mathcal{L} \supset \frac{1}{4\Lambda} \sum_{X} (C_{X})^{pq} (\mathcal{O}_{X})_{pq} + \text{H.c.}
\end{equation}
The factor of $1/4$ in Eq.~\eqref{eq:dim-5-lag} accounts for the permutation
symmetries of the operators. As suggested in Table~\ref{tab:eff-ops}, for each term there are $n_f (n_f + 1) / 2$
independent complex coefficients for $n_f$ flavours, since each coefficient
matrix is symmetric in flavour by Fermi--Dirac statistics. Di-quark couplings for the leptoquark $\chi^{d}$ are generated after
the breaking of $\mathrm{SU}(4) \otimes \mathrm{SU}(2)_{R}$:
\begin{equation}
  (\mathcal{O}_{X})_{pq} \supset \frac{4v_R}{\sqrt{2}} (d_{X})_{p}^{a} (u_{X})_{q}^{b} (\chi^{d})^{c} \epsilon_{abc} \ ,
\end{equation}
where $a,b,c$ are colour indices, thus allowing proton decay. Integrating out the field $\chi^d$ at tree level, the dimension-6 operators
\begin{align}
\label{eq:SLRudd}
\mathcal{O}_{udd}^{S,LR} &= \epsilon_{abc} (u^a_L d^b_L)(\bar{\nu}_L d^c_R) \ , \\
\label{eq:SRRudd}
\mathcal{O}_{udd}^{S,RR} &= \epsilon_{abc} (u^a_R d^b_R)(\bar{\nu}_L d^c_R) \ ,
\end{align}
are generated in the LEFT basis of
Ref.~\cite{Jenkins:2017jig}, where generational indices have been suppressed for clarity. Assuming one generation and using Eq.~\eqref{eq:physical-neutrinos}, we find the following $\Delta B = 1$ dimension-6 effective Lagrangian:
\begin{equation}
  \label{eq:dim-6-efflag}
  \mathcal{L}^{(6)} \supset - \sum_X \frac{ C_{X} y_{R} v_R \sin \theta }{\sqrt{2} \Lambda m_{\chi^{d}}^{2}} \mathcal{O}^{S,XR}_{udd} + \text{H.c.} \ ,
\end{equation}
at tree level, where $\sin \theta = m_{u} / m_{N} \approx m_{u} / |y_{R} v_{R}|$.

\section{Nucleon decay}
\label{sec:proton-decay}

\begin{figure}[t]
    \centering
    \includegraphics{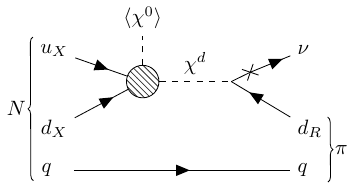}
    \caption{The diagram shows the nucleon decay induced by the effective operators $\mathcal{O}_{X}$, with $X\in \{L,R\}$. Here $q \in \{u,d\}$ and the cross represents the mixing between the singlet fermion $\xi$ and the light neutrino $\nu$.}
    \label{fig:decays}
\end{figure}

The operators $\mathcal{O}^{S,RR}_{udd}$ and $\mathcal{O}^{S,LR}_{udd}$ give rise to the nucleon decays
$n \to \pi^{0} \nu$ and $p \to \pi^{+} \nu$. We show these diagrammatically in Fig.~\ref{fig:decays} with the operators $\mathcal{O}^{S,XR}_{udd}$ resolved. Importantly, the contributions from these operators to the nucleon decays will add coherently\footnote{This opens up the possibility of a precise fine tuning to avoid these decays, but we neglect this possibility in the following discussion.}. The relevant rates can be estimated as
\begin{equation}
  \label{eq:one-gen-pion-decay}
  \Gamma(p \to \pi^{+} \nu) = \frac{m_{p}m_u^2}{16\pi \Lambda^2 m_{\chi^d}^4} \left| \sum_X C_{X} \langle \pi^{+} | (ud)_{X}d_{R} | p \rangle \right|^{2}
\end{equation}
and~\cite{Aoki:2013yxa}
\begin{equation}
  \Gamma(n \to \pi^{0} \nu) = \frac{1}{2} \Gamma(p \to \pi^{+} \nu) \ ,
\end{equation}
neglecting the mass difference between $p$ and $n$. A prediction of this scenario is thus the absence of nucleon decays with charged-lepton final states. Lattice calculations of the hadronic matrix elements give $\langle \pi^{+} | (ud)_{R}d_{R} | p \rangle = \SI{0.151}{\GeV^{2}}$ and $\langle \pi^{+} | (ud)_{L}d_{R} | p \rangle = \SI{-0.159}{\GeV^{2}}$~\cite{Yoo:2021gql}. The experimental limit on the neutron decay is somewhat stronger than that on the proton decay, with $\tau(n \to \pi^{0} \nu) > \SI{1.1e33}{\year}  $ at 90\% confidence~\cite{Super-Kamiokande:2013rwg}. We derive the following reference limit by turning on only $C_L$:
\begin{equation}
  \label{eq:breaking-scale-limit}
  m_{\chi^{d}} \gtrsim \SI{8e6}{\GeV}  \sqrt{|C_L|  \frac{\SI{e19}{\GeV}}{\Lambda} \frac{m_u}{\SI{171}{\GeV}}} \ .
\end{equation}
This, in combination with Eq.~\eqref{eq:mchid-vr-relation}, provides a direct bound on the $\mathrm{SU}(4) \otimes \mathrm{SU}(2)_R$ breaking scale:
\begin{equation}
  \label{eq:vR-bound}
  v_R \gtrsim \SI{5e6}{\GeV} \sqrt{|C_L|  \frac{\SI{e19}{\GeV}}{\Lambda}  \frac{m_u}{\SI{171}{\GeV}}  \frac{2\pi}{|\lambda^{\rchi}|}} \ .
\end{equation}
Importantly, the limit scales inversely to the square root of $\Lambda$, the scale of baryon-number violation. As a benchmark, we have taken this to be the Planck scale, where all global symmetries are expected to be violated~\cite{Hawking:1975vcx,Krauss:1988zc,Banks:2010zn,Harlow:2018tng}. In the following section, we show that it is in fact motivated to take $\Lambda$ to be the scale of $\mathrm{SO}(10)$ unification, which improves the limits by a factor of $\sqrt{\Lambda_{\text{Planck}}/\Lambda_{\text{GUT}}}$. Taking $\Lambda_{\text{GUT}} \sim \SI{e16}{\GeV}$ gives
\begin{equation}
  \label{eq:gut-bound}
  v_R \gtrsim \begin{pmatrix}
  \SI{4e5}{\GeV} \\
  \SI{9e6}{\GeV} \\
  \SI{1e8}{\GeV} 
  \end{pmatrix} \ ,
\end{equation}
where the three values represent the choices to set $m_u$ in Eq.~\eqref{eq:vR-bound} to the up, charm and top masses, and the other benchmark values are as before with $|C_L|=1$. The analogous limits for the Planck-scale suppressed operators are:
\begin{equation}
  \label{eq:planck-bound}
  v_R \gtrsim \begin{pmatrix}
  \SI{1e4}{\GeV} \\
  \SI{3e5}{\GeV} \\
  \SI{5e6}{\GeV} 
  \end{pmatrix} \ .
\end{equation}
In the following we motivate the presence of $m_c$ and $m_t$ in the limits by extending the one-generational model to three generations. However, we emphasis that even using the up-quark mass generates sizeable limits, comparable to the usual PS breaking limits. For the discussion below we also assume a UV scale of $\Lambda_\text{GUT}$.

In Eq.~\eqref{eq:breaking-scale-limit} we have intentionally expressed the limits on $m_{\chi^d}$ as a function of the dimensionless coupling constants $C_X$, since their values are unknown. Should these be small, the limits on the PS breaking scale could be significantly suppressed compared to those presented in Eqs.~\eqref{eq:gut-bound} and \eqref{eq:planck-bound}.
As an illustrative example, if the limits on $m_{\chi^d}$ are desired to be less than $1$~TeV, this requires:
 \begin{equation}
  \label{eq:CX-bound}
  |C_X| \lesssim \begin{pmatrix}
  \SI{1e-3}{} \\
  \SI{2e-6}{} \\
  \SI{2e-8}{} 
  \end{pmatrix}\frac{\Lambda}{10^{19}\text{ GeV}} \ ,
\end{equation}
again assuming up-, charm- and top-quark masses. Such incredibly small values for $C_X$ may occur in a UV theory in a technically natural way as $U(1)_B$ is recovered up to dimension 6 when $C_X \rightarrow 0$ in our minimal model, similar to the inverse-seesaw mechanism and the Weinberg operator.

\paragraph*{Three-generational mixing} Below we generalise to the complete three-generational model, which allows for a discussion of the effects of the quark--lepton mixing matrices, the full neutrino mixing, and opens up nucleon decays with kaons in the final state. We intend the following discussion to simply illustrate how the aforementioned higher-dimensional operators produce significant limits on the breaking scale of minimal Pati--Salam. We do not try to conduct an exhaustive scan of the parameter space to derive lower and upper bounds on the PS breaking scale from nucleon decay limits when different flavour structures are assumed in the theory.

Moving to three generations, the Lagrangian of Eq.~\eqref{eq:dim-6-efflag} now takes the form\footnote{In this expression, we have redefined $C_X$ to absorb any unphysical mixing matrices from the dimension-6 operators.}
\begin{equation}
  \mathcal{L}^{(6)} \supset - \sum \frac{(C_{X})_{pq} v_R}{\sqrt{2} \Lambda m_{\chi^{d}}^{2}} \Omega_{rs} (\mathcal{O}^{S,XR}_{udd})_{pqrs} + \text{H.c.} \ ,
\end{equation}
where the sum is over $X \in \{L,R\}$ and the flavour indices $p,q,r,s \in \{1,2,3\}$, which enumerate the generation of the fermions as they appear in Eqs.~\eqref{eq:SLRudd} and \eqref{eq:SRRudd}. As suggested by Eq.~\eqref{eq:one-gen-mixing},
the neutrino mixing is proportional to the up-quark mass matrix, and therefore the analogues of Eqs.~\eqref{eq:breaking-scale-limit} and \eqref{eq:vR-bound} are generically dominated by contributions proportional to $m_t$, up to specific textures for the fermion mixing matrices which we briefly explore below. The components $\Omega_{pq}$ are elements of the matrix
\begin{equation}
\label{eqn:3gendecaymixing}
  \boldsymbol{\Omega} \equiv \mathbf{K}_L^{u\nu \dagger} \mathbf{M}_u [v_R \mathbf{Y}_R^{\textsf{T}}]^{-1} \mathbf{Y}_R \mathbf{V}_{\text{CKM}}^R
\end{equation}
and they couple the physical $\nu_p$ to $d_{Rq}$ through $\chi^d$. The matrix $\mathbf{\Omega}$ is defined in terms of the mixing matrices introduced in Eq.~\eqref{eqn:mixingmatrixconnection}, $\mathbf{M}_u \equiv \mathrm{diag}(m_u, m_c, m_t)$ and the matrix of Yukawa couplings $\mathbf{Y}_R$ [the analogue of $y_R$ from Eq.~\eqref{eq:xi-lag}], defined in a mixed flavour-physical basis such that $\mathbf{Y}_{R} \mathbf{V}_{\text{CKM}}^R$ couples $\xi$ to $\chi^d$ and the physical down-type quarks. Note that Eq.~\eqref{eqn:3gendecaymixing} agrees with Eqs.~\eqref{eq:one-gen-mixing} and~\eqref{eq:dim-6-efflag} when one-generation is assumed.
\begin{widetext}

The generalisation of Eq.~\eqref{eq:one-gen-pion-decay} in the full model is
\begin{equation}
\label{eq:three-gen-pion-decay}
\Gamma(p \to \pi^+ \nu) = \frac{m_p v_R^2}{16\pi \Lambda^2 m_{\chi^d}^4} \left| \sum_{X,p} (C_X)_{11} \Omega_{p1} \langle \pi^+ | (ud)_X d_R | p \rangle  \right|^2 \ ,
\end{equation}
where the dependence of the expression on the up-type quark masses enters through Eq.~\eqref{eqn:3gendecaymixing}. In addition to the pion modes, the decays $p \to K^+ \nu$ and $n \to K^0 \nu$ are also relevant. The strange quark that should appear in the operators $\mathcal{O}^{S,LR}_{udd}$ and $\mathcal{O}^{S,RR}_{udd}$ to open up these decay channels can come about from the components $\Omega_{p2}$, which couple the light neutrinos to the strange quark through the leptoquark $\chi^d$, or else directly from the operators $(\mathcal{O}_{X})_{12}$ by turning Fig.~\ref{fig:decays} into a $t$-channel diagram:
\begin{equation}
\label{eq:three-gen-kaon-decay}
\Gamma(p \to K^+ \nu) = \frac{m_p v_R^2}{16\pi \Lambda^2 m_{\chi^d}^4} \left| \sum_{X,p} (C_X)_{12} \Omega_{p1} \langle K^+ | (us)_X d_R | p \rangle + \sum_{X,p} (C_X)_{11} \Omega_{p2} \langle K^+ | (ud)_X s_R | p \rangle \right|^2 \ .
\end{equation}
\end{widetext}
The hadronic matrix elements $\langle K^+ | (us)_L d_R | p \rangle = \SI{-0.0398}{\GeV^{2}}$ and $\langle K^+ | (us)_R d_R | p \rangle = \SI{0.0284}{\GeV^{2}}$ are suppressed with respect to $\langle K^+ | (ud)_L s_R | p \rangle = \SI{-0.109}{\GeV^{2}}$ and $\langle K^+ | (ud)_R s_R | p \rangle = \SI{0.1006}{\GeV^{2}}$~\cite{Yoo:2021gql}. Here the experimental limits on the proton decay are two orders of magnitude more stringent than those on the corresponding neutron decay: $\tau(p \to K^+ \nu) \geq \SI{5.9e33}{\year}$ at 90\% confidence~\cite{Super-Kamiokande:2014otb}. We derive reference limits on $v_R$ for the decay $\Gamma(p \to K^+ \nu)$, as we did earlier for the pionic decays. Turning on only $C_L$ and assuming diagonal textures for the matrices in Eq.~\eqref{eqn:3gendecaymixing}, we find
\begin{equation}
 v_R \gtrsim \SI{1.3e7}{\GeV} \ .
\end{equation}
The choice to set the mixing matrices diagonal in this case implies $\mathbf{\Omega} = \mathbf{M}_u$, and therefore $\Gamma(p \to K^+ \nu) \propto m_c^2$. Taking more democratic textures yields $\Gamma(p \to K^+ \nu) \propto m_t^2$, and limits of order $\SI{e8}{\GeV}$ for $\mathcal{O}(1)$ coefficients. 

The above discussion highlights the importance of the mixing-matrix textures that enter in Eq.~\eqref{eqn:3gendecaymixing}. Importantly, low-scale PS scenarios already require special textures for $\mathbf{K}_{L/R}^{d e}$ in order to suppress FCNC $K_L^0$ decays~\cite{Kuznetsov:1994tt,Valencia:1994cj,Kuznetsov:1995wb,Smirnov:2007hv,Smirnov:2008zzb,Smirnov:2018ske,Dolan:2020doe} mediated by $X_\mu$. As both $\mathbf{V}_{\text{CKM}}^L $ and $\mathbf{U}_{\text{PMNS}}^L$ are fixed in the SM, fixing the structure of $\mathbf{K}_{L/R}^{d e}$ to minimise the limits on $m_X$ has the unavoidable consequence of completely determining the structure of $\mathbf{K}_{L}^{u \nu}$ as can be seen in Eq.~\eqref{eqn:mixingmatrixconnection}. Of course, the right-handed unitary mixing matrices remain totally unconstrained.

As an instructive example\footnote{Here we assume $\mathbf{Y}_R$ is symmetric to simplify our discussion related to Eq.~\eqref{eqn:3gendecaymixing}. Its inclusion can only make it more difficult to construct flavour structures which prevent $m_t$ from dominantly contributing to the nucleon decays.}, we fix the structure of $\mathbf{K}_L^{de}$, and therefore $\mathbf{K}_{L}^{u \nu}$, using Table~4 of Ref.~\cite{Dolan:2020doe} such that the rare-meson decay limits induced by $X_\mu$ are fixed to one of their lowest possible values in minimal Pati--Salam: $\SI{81}{\TeV}$. In this case we find the flavour structure which enters the nucleon decay widths is roughly proportional to
\begin{equation}
    v_R \sum_j \Omega_{ji} \simeq 0.5\,m_u  \tilde{\mathbf{V}}_{1i} + 1.3\,m_c \tilde{\mathbf{V}}_{2i} + 1.0\,m_t \tilde{\mathbf{V}}_{3i}
\end{equation}
where $\tilde{\mathbf{V}}$ is the matrix resulting from absorbing the overall phase from each term into the rows of $\mathbf{V}_{\text{CKM}}^R$. To completely suppress the top-quark contribution requires that the $(3,1)$ and $(3,2)$ entries of $\tilde{\mathbf{V}}$ be zero\footnote{More realistically, this requires these two entries to be less than about $\mathcal{O}(10^{-3})$, such that the top mass is suppressed over the charm mass. For reference, in the SM $(\mathbf{V}^L_{\text{CKM}})_{32}\simeq 10^{-2}$.}, however the unitarity of the matrix now necessarily implies that $\lvert \tilde{\mathbf{V}}_{21}\rvert = \sin\theta$ and $\lvert \tilde{\mathbf{V}}_{22}\rvert = \cos\theta$, preventing a suppression of the charm- and up-mass contribution. Therefore, in the worst case scenario, the contribution proportional to the charm mass will dominate the nucleon decay, constraining $v_R$ to roughly $\mathcal{O}(10^7)$~GeV, significantly larger than the $\SI{81}{\TeV}$ implied by the $X_\mu$-mediated FCNC meson decays. 

Reversing the situation, we can fix the flavour structures of $\mathbf{V} = \mathbf{V}_{\text{CKM}}^R$ and $\mathbf{K} = \mathbf{K}_{L}^{u\nu}$ such that the relevant entries of $\mathbf{\Omega}$ are as suppressed as possible. Firstly, to suppress the top-mass contribution again requires the condition $\mathbf{V}_{31} = \mathbf{V}_{32} = 0$. One can also suppress the charm-quark contribution provided that the curious condition  
\begin{equation}
\label{eqn:killcharmquarkcontributioninit}
    \sum_i \mathbf{K}_{i2} = 0
\end{equation}
is satisfied.

The simplest way to satisfy such a constraint for a column of a unitary matrix is when one entry of the column is zero and the other two are $\pm 1/\sqrt{2}$ (with appropriate choices for the other columns of the matrix). This implies that the up-quark mass will be the only contribution to Eqs.~\eqref{eq:three-gen-pion-decay} and~\eqref{eq:three-gen-kaon-decay} and a limit of around $\mathcal{O}(100)$ TeV will be generated from nucleon decays. However, using Eq.~\eqref{eqn:mixingmatrixconnection} but now solving for $\mathbf{K}_L^{d e}$ we find that, for the limited possible choices of $\mathbf{K}_L^{u \nu}$ that satisfy Eq.~\eqref{eqn:killcharmquarkcontributioninit} with one entry zero, $\mathcal{O}(1)$ entries are generated in the upper-left $2\times 2$ block of $\mathbf{K}_L^{de}$. This implies that $X_\mu$ will receive stringent limits from $K_L^0$ decays of order $\mathcal{O}(1000)$ TeV~\cite{Dolan:2020doe}, which we have confirmed numerically assuming that $\mathbf{K}_R^{de}$ has a structure such that the $K_L^0$ decays are maximally suppressed. Numerically we also find other examples of unitary matrices which satisfy Eq.~\eqref{eqn:killcharmquarkcontributioninit}, where now each entry in the second column is nonzero, and unsurprisingly similar limits from $K_L^0$ decays arise. Therefore, engineering the flavour structure of the model such that these UV-induced nucleon decays are maximally suppressed appears to generate close to maximal limits from the usual $X_\mu$ mediated rare-meson decays. 

We find, in minimal Pati--Salam, that while specific flavour structures in the mixing matrices can suppress the limits on $m_X$ in rare meson decays, these structures generate even larger limits on $v_R$ (and therefore $m_X$) from $B$-violating nucleon decays. This is under the reasonable assumption that above the PS scale there exists UV physics generating the operators of Eq.~\eqref{eq:b-viol-operators}.

Before continuing we also comment briefly that the scenario we present here displays some general features that may help distinguish it from other models that predict nucleon decays.  First, as briefly touched on earlier, our model predicts that nucleon decays to neutrinos should dominate over decays to charged leptons, since the Yukawa couplings of the $\chi^d$ to charged leptons are absent at dimension 4. We can also expect a complementarity between nucleon-decay signals and other flavour observables, suppressed by powers of $v_R$. Since the diquark couplings of $\chi^d$ are generated at dimension 5, its flavour phenomenology is dominated by the leptoquark couplings. These mediate rare semileptonic decays of $B$ and $K$ mesons at tree level, \textit{i.e.} $B \to K^{(*)} \nu\nu$ and $K \to \pi \nu \nu$, and allow mixing effects in the neutral $B$ and kaon systems. Additionally, the vector leptoquark $X_\mu$, which should exist at a similar mass scale to $\chi^d$, allows FCNC processes such as the $K^0_L$ decays discussed above. The combined observation of signals arising from both $\chi^d$ and $X_\mu$ is therefore a key prediction in the theory. In contrast to many GUT scanarios, this setup thus predicts a rich flavour phenomenology, which can help pin down the model in the event that the nucleon decays to neutrino final states are observed.

\section{UV completions}

The operators $\mathcal{O}_{L}$ and $\mathcal{O}_{R}$ can arise via the tree-level exchange of field content present at scales
$\Lambda > v_R$. Below we derive the PS representations that are
relevant, using methods analogous to those used to systematically derive
tree-level completions for operators in the SMEFT~\cite{Gargalionis:2020xvt}. We aim to show that the tree-level generation of the operators $\mathcal{O}_{X}$ may be unavoidable in any sensible $\mathrm{SO}(10)$ embedding of our minimal setup.

There are two ways\footnote{We highlight the similarity here to the minimal, tree-level completions of the Weinberg operator, which has the same general structure as the $\mathcal{O}_{X}$.} a heavy multiplet could couple to $f_{X}$ and $\chi$ so as to generate at least one of the operators $\mathcal{O}_{L}$ or $\mathcal{O}_{R}$ through renormalisable interactions: (1) as a Lorentz scalar $S$ with couplings
\begin{equation}
\label{eq:S-lag}
    \mathcal{L} \supset - y_{X} S f_{X} f_{X} - \kappa S^{\dagger} \chi \chi \ ,
\end{equation}
or (2) as a Majorana fermion $F$ coupling as
\begin{equation}
\label{eq:F-lag}
    \mathcal{L} \supset -  z_{X} F f_{X} \chi 
 \end{equation}
where multiple singlet contractions may exist in each case and $F f_X$ is appropriately constructed to be a Lorentz scalar. It is important to note that the $\mathrm{SU}(4)$ representations of the heavy multiplets should be antisymmetric in order to generate a structure like Eq.~\eqref{eq:b-viol-operators}.

The $\kappa$ term imposes that the scalar $S$ form a singlet with
\begin{equation}
  \label{eq:chichi-decomposition}
  \begin{split}
    (\mathbf{4}, \mathbf{1}, \mathbf{2}) \otimes (\mathbf{4}, \mathbf{1}, \mathbf{2}) &= (\mathbf{10}, \mathbf{1}, \mathbf{3}) \oplus (\mathbf{6}, \mathbf{1}, \mathbf{1}) \\
                                                                                      &\quad \oplus (\mathbf{10}, \mathbf{1}, \mathbf{1}) \oplus (\mathbf{6}, \mathbf{1}, \mathbf{3}) \ .
  \end{split}
\end{equation}
The $\mathbf{10}$ is symmetric, and so the first and third representations are discounted. Similarly, the triplet representations under $\mathrm{SU}(2)_{R}$ cause the $\kappa$ term to vanish identically for a single $\chi$ generation. This fixes the assignment $S \sim (\mathbf{6}, \mathbf{1}, \mathbf{1})$ as the only option. This multiplet generates both $\mathcal{O}_{L}$ and $\mathcal{O}_{R}$.
  
The fermion $F$ could couple to $f_R$ or $f_L$. In the former case, the allowed representations are the $(\mathbf{6}, \mathbf{1}, \mathbf{1})$ and $(\mathbf{6}, \mathbf{1}, \mathbf{3})$, both of which are viable and generate only $\mathcal{O}_{R}$. In the latter case, $F$ should form a singlet with
\begin{equation}
  \label{eq:LR-decomposition}
  \begin{split}
    (\mathbf{4}, \mathbf{2}, \mathbf{1}) \otimes (\mathbf{4}, \mathbf{1}, \mathbf{2}) &= (\mathbf{10}, \mathbf{2}, \mathbf{2}) \oplus (\mathbf{6}, \mathbf{2}, \mathbf{2}) \ ,
  \end{split}
\end{equation}
of which only the $(\mathbf{6}, \mathbf{2}, \mathbf{2})$ works, again due to its antisymmetric $\mathrm{SU}(4)$ indices. This multiplet only generates $\mathcal{O}_{L}$ at tree level.

\paragraph*{Embedding into $\mathrm{SO}(10)$} It is usual to imagine that the PS gauge group is embedded into $\mathrm{SO}(10)$, of which it is a maximal subgroup, at the scale $\Lambda_{\mathrm{GUT}} \sim \SI{e16}{\GeV}$. The PS multiplet $\Phi \sim (\mathbf{1}, \mathbf{2}, \mathbf{2})$, if it is to couple to the SM fermions as in Eq.~\eqref{eq:ps-yukawas}, must be contained within 
\begin{equation}
  \mathbf{16} \otimes \mathbf{16} = \mathbf{10}_{S} \oplus \mathbf{120}_{A} \oplus \overline{\mathbf{126}}_{S} \ .
\end{equation}
The associated branching rules are
\begin{equation}
\begin{split}
    \mathbf{10} &\rightarrow \mathbf{(6,1,1) \oplus (1,2,2)} \ , \\
    \mathbf{120} &\rightarrow \mathbf{(1,2,2)} \oplus \mathbf{(10,1,1)} \oplus (\overline{\mathbf{10}},\mathbf{1},\mathbf{1})\oplus \mathbf{(6,1,3)} \\
    &\qquad \oplus \mathbf{(6,3,1)}\oplus \mathbf{(15,2,2)} \ , \\
    \mathbf{126} &\rightarrow \mathbf{(6,1,1) \oplus (10,1,3) \oplus (10,3,1) \oplus (15,2,2)} \ ,
\end{split}
\end{equation}  
and we highlight that only the $\mathbf{120}$, owing to its antisymmetry, does not contain the field $S$, which generates both $\mathcal{O}_{L}$ and $\mathcal{O}_{R}$. Thus, any embedding of the described low-scale PS model into $\mathrm{SO}(10)$ should place the bidoublet $\Phi$ into only the 120-dimensional representation to avoid the tree-level generation of $\mathcal{O}_{L}$ and $\mathcal{O}_{R}$. This is problematic, since the antisymmetry of the $\mathbf{120}$ impresses an antisymmetric flavour structure onto the SM Yukawa couplings. Such a model implies the equality of the masses of two fermion generations to all orders at the scale $\Lambda$, in the absence of some flavour-specific dynamics. Thus we conclude that the $\mathbf{10}$ or $\overline{\mathbf{126}}$ are necessary elements of a realistic $\mathrm{SO}(10)$ theory, and therefore  the operators $\mathcal{O}_L$ and $\mathcal{O}_R$ are unavoidably generated at the GUT scale.

\section{Viability of a light $X_\mu$}

Below we discuss modifications to the minimal low-scale theory that may be able to accommodate a light $X_\mu$ gauge boson, while still being consistent with the nucleon decay bounds we have presented above. We also comment briefly on the extent to which the arguments presented so far can be applied to popular variants of low-scale PS in the literature that can naturally arrange for a light $X_\mu$ gauge boson.
 
First, we emphasise that the bound derived in Eq.~\eqref{eq:breaking-scale-limit} is particular to the case of the $\mathrm{SU}(4) \otimes \mathrm{SU}(2)_{L} \otimes \mathrm{SU}(2)_{R}$ gauge group. The operators $\mathcal{O}_{L}$ and $\mathcal{O}_{R}$ depend on the balance of antisymmetry coming from their $\mathrm{SU}(4)$ and $\mathrm{SU}(2)_R$ structures, and they can be made to vanish identically by demoting $\mathrm{SU}(2)_{R}$ instead to a $\mathrm{U}(1)_R$ symmetry, again, for a single generation of $\chi$. This Pati--Salam-like gauge group could still be consistent with $\mathrm{SO}(10)$ unification if, for example, one imagines that $\mathrm{SU}(2)_{R}$ is broken to its $\mathrm{U}(1)_R$ subgroup at some very large scale owing to the fact that the PS breaking scalar necessarily is charged under $\mathrm{SU}(2)_R$. This is easily achieved with the vev of a heavy $\mathrm{SU}(2)_{R}$-triplet scalar: $(\mathbf{1},\mathbf{1},\mathbf{3})$. In such a scenario the scalar $\chi$ breaks up into two $\mathrm{SU}(4)$-fundamental scalars, one of which is appointed to break $\mathrm{SU}(4)$ at a much lower scale, while the other naturally lives at the scale of $\mathrm{SU}(2)_R$ breaking. This latter, heavy scalar contains the $\chi^d$ and therefore the nucleon decay could be kept under control, but a detailed analysis of such a model is beyond the scope of this paper. We note that such a scenario would imply that the $W'_\mu$ gauge boson cannot live at a common (low) scale with the $X_\mu$ and $Z'_\mu$, that is $m_{W'} \gg m_{Z'},\,m_X$.

We also point out that the $(\mathbf{4}, \mathbf{1}, \mathbf{2})$ is not the only possible representation for the PS-breaking scalar, and the conclusions of this paper do not immediately apply to alternative choices. PS-breaking scalars must have the appropriate quantum numbers to preserve the $U(1)_Y$ of the SM. The next largest-dimensional scalar commonly employed in PS model building after the $(\mathbf{4}, \mathbf{1}, \mathbf{2})$ is the $(\mathbf{10},\mathbf{1},\mathbf{3})$. This scalar generates no dimension-5 operators whatsoever, assuming standard fermion embeddings. While this may seem like a good candidate to avoid unwanted nucleon decays, such a scalar cannot allow for light neutrino masses without $v_R \gtrsim \mathcal{O}(10^{12})$~GeV, if a seesaw explanation is desired, preventing light leptoquarks from appearing in the theory. It may be an interesting model-building direction to try and generate light neutrino masses without tuning by employing this scalar. Of course, it is possible that whatever field content is introduced may end up generating dangerous dimension-5 $B$-violating operators. 

To the best of our knowledge higher-dimensional scalar representations beyond these two have not been used in any models of high- or low-scale PS, as such scalars will not couple to the SM fermions in a renomarmalisable way. In such cases, the $m_u = m_\nu$ prediction of PS can only be broken through extreme tuning between multiple EW Higgs multiplets. Introducing exotic fermions of appropriate dimension to couple such a PS-breaking scalar to the SM fermions may avoid this tuning and allow for low-scale PS breaking, but this must be evaluated on a model-by-model basis before analysing the implications of dimension-5 $B$-violating operators. The introduction of large-dimensional fermion multiplets will inevitably generate extra mass mixing for the SM fermions, \textit{e.g.} charged-lepton mass mixing, and therefore additional nucleon decay channels will generically be predicted beyond the neutral channel appearing in the minimal model if $B$ violation occurs at dimension 5.

\paragraph*{Modified Pati--Salam theories} Models based on the ``Pati--Salam-adjacent'' gauge group\footnote{A similar variant first appears in the original paper~\cite{Pati:1974yy} and was referred to as the ``economical'' model.} $\mathrm{SU}(4) \otimes \mathrm{SU}(3) \otimes \mathrm{SU}(2) \otimes \mathrm{U}(1)$ have gained popularity recently as useful frameworks for arranging for an ultra-light $X_\mu$ gauge boson, while at the same time avoiding the stringent meson-decay limits that the minimal model suffers from. Such a gauge structure can arise from the breaking of multiple copies of the Pati-Salam gauge group~\cite{Bordone:2017bld,FernandezNavarro:2022gst}. These models are successful in explaining the flavour anomalies in charged- and neutral-current $B$-meson decays. We take~\cite{DiLuzio:2017vat} as an illustrative example of such a model and find that even here, in a model without an $\mathrm{SU}(2)_R$ gauge group, a dimension-5 $B$-violating operator can also be written down in analogy to the operators $\mathcal{O}_X$:
\begin{equation}
    \mathcal{L}^{(5)} \supset C d_R d_R \Omega_1^\dagger \Omega_3 \ ,
\end{equation}
where $\Omega_1 \sim (\overline{\mathbf{4}},\mathbf{1},\mathbf{1},-1/2),\,\Omega_3 \sim (\overline{\mathbf{4}},\mathbf{3},\mathbf{1},1/6)$, $d_R \sim (\mathbf{1},\mathbf{3},\mathbf{1},-1/3)$, $C$ is the operator coefficient, and flavour indices are implicit. We note that $d_R d_R$ must be antisymmetric in flavour. The scalar $\Omega_3$ is required to break $\mathrm{SU}(4) \otimes \mathrm{SU}(3) \rightarrow \mathrm{SU}(3)_c$, while $\Omega_1$ is required for phenomenological reasons, see \textit{e.g.}~\cite{DiLuzio:2018zxy, Greljo:2018tuh}. The operator generates di-quark couplings for the scalar leptoquarks in the theory in a similar way. The fields $\Omega_1$, $\Omega_3$ and $d_R$ seem to appear consistently for all model variants of this gauge group, and therefore nucleon decay limits may be a concern\footnote{We note that the nucleon decays in this model will be suppressed by loop and CKM factors due to the antisymmetry in $d_R d_R$.} if one imagines such a theory couples to UV physics that generates Eq.~\eqref{eq:b-viol-operators}. This is particularly true in this instance, as the primary goal in such models is to organise for a TeV scale $X_\mu$ leptoquark. It may be the case that modifications of the theory can be made to avoid these effects; for example, without fundamental scalars in the theory~\cite{Fuentes-Martin:2020bnh} the dangerous operator may be avoided entirely.

A simpler alternative to modifying the gauge structure of the theory is to instead introduce additional fermionic multiplets which, when broken to the SM, contain states that will mix with the down quark or charged-lepton states. Such models can cause $X_\mu$ to couple dominantly to one chirality of $d$ and $e$ (\textit{e.g.} $\overline{d_L} \slashed{X} e_L$) but not the other, such that a ``Chiral Pati--Salam'' model is achieved\footnote{Again, this idea dates back to the original paper~\cite{Pati:1974yy} where it was referred to as the ``prodigal'' model.}. This is possible by engineering the fermion multiplets and mass mixing matrices such that the physical $d$ and $e$ states do not arise from the same $\mathrm{SU}(4)$ multiplet~\cite{Foot:1997pb,Foot:1999wv,Balaji:2018zna,Balaji:2019kwe,Dolan:2020doe,Iguro:2021kdw}, suppressing the $X_\mu$ couplings to these states for one specific chirality. We note that this can also simultaneously break the down-isospin mass degeneracy predicted by PS~\cite{Dolan:2020doe} so can be a simple, attractive and possibly testable PS variant. This induces a helicity suppression in the rare-meson decays in analogy with the helicity suppression of pion decays in the SM. Such a scenario shares a common gauge structure and scalar content with the minimal model discussed in this paper, and therefore  suffers from similar UV-induced $B$-violating nucleon decays. However, while an estimate of the constraints on the $\chi^d$ mass from nucleon decay will be model dependent, we expect that here the limits on $m_{\chi^d}$ will be even larger. In this case, the light leptons are usually assumed to exist outside of the PS multiplets $f_X$, but are necessarily Yukawa coupled to them with $\chi$ in order to generate the required mass mixing. As a result, the $\sin \theta$ suppression of Eq.~\eqref{eq:dim-6-efflag} will likely be absent, and therefore much larger decay rates will follow, requiring an even larger value of $m_{\chi^d}$ in order to suppress them.

In both variants of Pati--Salam discussed above, exotic fermions which mix with the SM particle content are predicted. As a result, in both cases, additional nucleon decay channels may open up, including decays with charged-lepton final states, which often have even larger bounds compared to the neutrino channels predicted in minimal Pati--Salam. The nucleon decay modes that dominate can vary between models, and therefore we believe it is important to estimate the effects of possible dimension-5 $B$-violating operators on each variant of Pati--Salam introduced to allow for lower limits on $X_\mu$.

\section{Conclusions}

In this paper we have studied the effects of dimension-5 $B$-violating operators in the context of the minimal, low-scale Pati--Salam model. Our results show that the nucleon decays mediated by these operators can lead to significant constraints on the mass of the scalar leptoquark $\chi^d$, and therefore the scale of $\mathrm{SU}(4) \otimes \mathrm{SU}(2)_{R}$ breaking. The operators are necessarily present in any reasonable $\mathrm{SO}(10)$ embedding of the model, and therefore the lower-bound on the breaking scale could be as large as $\mathcal{O}(10^8)$~GeV, under reasonable parameter choices. We point out that even if they are suppressed by the Planck scale, they can lead to unacceptably large nucleon-decay rates. 

Attempts to suppress the problematic nucleon decays push the model into a region of parameter space that implies dangerously large rates for the well known FCNC meson decays mediated by the $\mathrm{SU}(4)$ gauge-boson leptoquark. Our study highlights the importance of considering EFT scenarios beyond the SM, especially at low mass dimension. We leave open the possibility of building a low-scale Pati--Salam model with an accidental $\mathrm{U}(1)_B$ that remains unbroken at dimension-5, while naturally generating small neutrino masses. The extent to which this phenomenon can generalise beyond the minimal model we have presented here is also an interesting line of future work. This may be particularly relevant for known Pati--Salam variants designed to allow for a TeV-scale $X_\mu$ leptoquark.

\section*{Acknowledgements}

We thank Admir Greljo and Raymond R. Volkas for useful discussions and comments on the manuscript. We also thank Peter Stangl for useful correspondence. TPD would like to thank Arcadi Santamar\'{i}a and remaining members of the Department of Theoretical Physics at the University of Valencia for their kind hospitality while some of this work was performed. Feynman diagrams were
generated using the Ti\textit{k}Z-Feynman package for
\LaTeX~\cite{Ellis:2016jkw}. We also acknowledge the use of \textsf{Sym2Int} for
group-theory calculations~\cite{Fonseca:2019yya, Fonseca:2017lem}. TPD is supported by KIAS Individual Grants under Grant
No. PG084101 at the Korea Institute for Advanced Study. JG is supported by the MICINN/AEI  (10.13039/501100011033) grant  PID2020-113334GB-I00 and the ``Generalitat Valenciana" grants PROMETEO/2021/083 and PROMETEO/2019/087.

\bibliography{main}

\end{document}